\documentclass[manuscript]{acmart}
\acmConference[ICSE 2024]{46th International Conference on Software Engineering}{April 2024}{Lisbon, Portugal}

\AtBeginDocument{%
  \providecommand\BibTeX{{%
    \normalfont B\kern-0.5em{\scshape i\kern-0.25em b}\kern-0.8em\TeX}}}

\setcopyright{acmcopyright}
\copyrightyear{2024}
\acmYear{2024}
\acmDOI{XXXXXXX.XXXXXXX}

\acmBooktitle{ICSE 2024: International Conference on Software Engineering 2024}
\acmPrice{15.00}
\acmISBN{978-1-4503-XXXX-X/18/06}

\usepackage{eiffel}
\usepackage{soul}
\usepackage{enumitem}
\setitemize{leftmargin=*}

\newcommand{\e}[1]{\mbox{\lstinline[basicstyle=\normalsize]|#1|}}

\newcommand{\foote}[1]{\mbox{\color[HTML]{222277}\footnotesize\bfseries\ttfamily}}

\copyrightyear{2024}
\acmYear{2024}
\setcopyright{acmlicensed}\acmConference[APR '24 ]{2024 ACM/IEEE International Workshop on Automated Program Repair}{April 20, 2024}{Lisbon, Portugal}
\acmBooktitle{2024 ACM/IEEE International Workshop on Automated Program Repair (APR '24 ), April 20, 2024, Lisbon, Portugal}
\acmDOI{10.1145/3643788.3648007}
\acmISBN{979-8-4007-0577-9/24/04}

\begin{document}

%%
%% The "title" command has an optional parameter,
%% allowing the author to define a "short title" to be used in page headers.
    \title{BUGFIX: towards a common language and framework for the Automatic Program Repair community}

%%
%% The "author" command and its associated commands are used to define
%% the authors and their affiliations.
%% Of note is the shared affiliation of the first two authors, and the
%% "authornote" and "authornotemark" commands
%% used to denote shared contribution to the research.
\author{Bertrand Meyer}
\email{Bertrand.Meyer@inf.ethz.ch}
\orcid{0000-0002-5985-7434}

\author{Viktoryia Kananchuk}
\email{viktoryia.kononchuk@sit.study}
\orcid{0009-0000-3008-6245}

\author{Li Huang}
\email{Li.Huang@constructor.org}
\orcid{0000-0003-3531-4045}

\affiliation{%
  \institution{Constructor Institute}
  \streetaddress{Rheinweg 9}
  \city{Schaffhausen}
  \country{Switzerland}
  \postcode{8212}
}

\newcommand{\toolname}{Proof\-2\-Fix}

%%
%% By default, the full list of authors will be used in the page
%% headers. Often, this list is too long, and will overlap
%% other information printed in the page headers. This command allows
%% the author to define a more concise list
%% of authors' names for this purpose.

%%
%% The abstract is a short summary of the work to be presented in the
%% article.
\begin{abstract}
Techniques of Automatic Program Repair (APR) have the potential of thoroughly facilitating the task of producing quality software. After a promising start, however, progress in making APR practical has been hindered by the lack of a common framework to support the multiplicity of APR ideas and tools, and of target programming languages and environments. In this position paper we outline a general framework to enable the APR community to benefit from each other’s  advances, in particular through a standard language for describing bugs and their fixes. Such a common framework --- which is also applicable to work on fault seeding --- could be a tremendous benefit to researchers and developers of Interactive Development Environments (IDEs) who are working to make APR an effective part of the software developer’s practical experience.
\end {abstract}

%%
%% The code below is generated by the tool at http://dl.acm.org/ccs.cfm.
%% Please copy and paste the code instead of the example below.
%%


%%
%% Keywords. The author(s) should pick words that accurately describe
%% the work being presented. Separate the keywords with commas.

%% A "teaser" image appears between the author and affiliation
%% information and the body of the document, and typically spans the
%% page.

%%
%% This command processes the author and affiliation and title
%% information and builds the first part of the formatted document.
\maketitle

\section{Introduction} \label{introduction}

 We expect software to work properly; failures to meet that expectation, known as faults or bugs, can have serious consequences, all the more serious that software is now pervasive in the pursuit of almost all human affairs. Yet, just as the best-behaved children will occasionally do something silly and the most careful drivers will occasionally get a speeding ticket, programmers will occasionally produce buggy software. Not just human programmers but automated ones as well: Artificial Intelligence assistants such as ChatGPT and Copilot, while impressive in their capabilities, cannot be trusted to produce bug-free software; they mess up just as much as we, mere mortals \cite{meyer2023ai}. %[REFERENCE TO MY CACM BLOG, MAKE IT CLEAR THAT IT IS A BLOG ARTICLE].

Techniques of software verification, from tests to static analysis and proofs, help identify bugs; in recent years the idea has emerged that while spotting a bug is good, correcting the bug --- or, more realistically, suggesting a correction to the programmer --- is better. Some early work in this direction includes \cite{liu2018lsrepair, nguyen2013semfix, 7463060, 6776507, 7203042, 10.1145/3180155.3180250}; good surveys can be found in \cite{monperrus2018automatic, 
monperrus2018living, gazzola2018automatic}. Ideally, Automatic Program Repair (APR) should be included in the basic toolset that programmers use (IDE, Interactive Development Environment), so that if a programmer produces a potentially faulty code element, the IDE’s verification tools silently detect it and pop up a message that both signals the bug and suggests one or more valid corrections.
Making this scenario possible is the common goal --- the common Graal --- of all developers and researchers working on APR. One of the major obstacles to  reaching it is the heterogeneity of the field. Heterogeneity of APR approaches; diversity of IDEs; diversity of programming languages; and also diversity of bugs. These heterogeneity factors force every APR project to invest a major part of its effort in recreating a basic bug and fix description framework; such spurious repetition of work considerably impair progress towards making Automatic Program Repair a standard and effective part of the software developer’s daily experience.
The goal of the work-in-progress described here is to overcome these obstacles by establishing a joint framework, BUGFIX, that all APR efforts can use. The present paper is explicitly intended as a workshop position paper, intended for discussion and feedback rather than presenting a fully developed solution. Our intent is to present BUGFIX at the workshop to elicit feedback from the APR community and go on to develop a finalized version with the best chances of being widely adopted, and as a result speeding up the development of Automatic Program Repair.

\section{A Bug-and-Fix specification language} \label{language}
At the core of BUGFIX lies a language for describing identified patterns for both bugs and their fixes. Two observations are in order:
\begin{itemize}
    \item The BUGFIX effort focuses on bugs that manifest themselves through code patterns that are wrong and should be replaced (fixed) by adapting the patterns. For example, code that uses \e{f (a, b)} when it should use \e{f (b, a)}. We do not at this point consider deeper or more elaborate bugs, such as design bugs, as analyzed for example in \cite{catolino2019not}.
    \item To illustrate the BUGFIX language, we use a concrete syntax. For clarity the syntax is keyword-oriented and Eiffel-like. Concrete syntax details are not important, however, for the concepts discussed here; the syntax may change in the future. What matters is the abstract syntax and underlying semantic concepts. In many practical applications we expect that BUGFIX will be used not through a human-oriented syntax such as the one illustrated below but through a program-oriented API (Abstract Program Interface), for use by APR tools and databases.
\end{itemize}

The following example, using an ad hoc concrete syntax as just mentioned, illustrates the argument-reversal bug and its fix.

BUGFIX should support a wide range of programming languages. It needs, however, to provide bug and fix descriptions for language mechanisms, or “constructs”,  that exist (in different forms) in various languages. Examples include routine call, assignment, loop and so on. The specification of language constructs is the first part of BUGFIX. 

The mechanism supports both the description of a general pattern for each construct and the specification of its realization in a particular language; the latter should be extendible, so that one can add new languages and their implementation of predefined general constructs. The general specification of “routine call” could correspondingly appear as
\begin{lstlisting}[captionpos=b, basicstyle=\fontsize{0.27cm}{0.27cm}]
    construct CALL feature
        args: EXPRESSION$^*$
        r: ROUTINE
    end
\end{lstlisting}
where the constructs \e{EXPRESSION} and \e{ROUTINE} are separately defined. The specification of \e{CALL} simply state that a routine call includes a routine name (\e{r}) and a list of actual arguments (\e{args}), each of which is an instance of \e{EXPRESSION}. The specific specifications syntax (in Java and Eiffel) are below:
\begin{figure}
\begin{lstlisting}[captionpos=b, basicstyle=\fontsize{0.27cm}{0.27cm}]
    syntax CALL for Java:
        r (args)
    syntax CALL for Eiffel:
        [args.count $\neq$ 0 $\rightarrow$ r (args) | r]
\end{lstlisting}
\end{figure}

Note the conditional expression in the Eiffel case: a call with actual arguments uses parentheses, as in “\e{r (a, b, c)}”, but a call without arguments is  (unlike in Java) just “\e{r}” without parentheses. BUGFIX includes support for such conditional mechanisms.

The separation between a general language construct, such as \e{CALL}, and the specification of its (concrete) syntax for any particular programming language, contributes to the generality of BUGFIX, its usability by many different APR projects targeting at different languages and IDEs, and its extendibility to new environments.

Once the language constructs have been identified, we may proceed to the core goal of BUGFIX --- specifying bug-and-fix patterns. Here is the swapped-argument example:
\begin{figure}[htbp]
\begin{lstlisting}[captionpos=b, basicstyle=\fontsize{0.27cm}{0.27cm}]    
    pattern SWAPPED_ARGUMENTS for
        c: CALL
    with
        a1, a2: EXPRESSION
    where
        a1 $\in$ c.args ; a2 $\in$ c.args
        a1.index $\neq$ a2.index
    fix
        c [a1 $\leftarrow$ a2, a2 $\leftarrow$ old a1] 
    end
\end{lstlisting}
\end{figure}
%        c [a1 $\leftarrow$ a2, a2 $\leftarrow$ a1]

\noindent The intent should be clear: we describe a program transformation whereby two out-of-order arguments are swapped. 

The following example covers two commonly encountered bugs:
a programmer uses a sum instead of a difference  (\e{PLUS_MINUS}), or use of equality operator instead of inequality (\e{EQ_NEQ}).
\begin{figure}[htbp]
\noindent\begin{minipage}{.45\textwidth}
\begin{lstlisting}[captionpos=b, basicstyle=\fontsize{0.27cm}{0.27cm}]  
    pattern PLUS_MINUS for
        e: SUM
    with
        e1, e2: EXPRESSION
    where
        e1 = e.first
        e2 = e.second
    fix
        DIFFERENCE [first $\leftarrow$ e1, second $\leftarrow$ e2]
    end
\end{lstlisting}
\end{minipage}
\noindent\begin{minipage}{.45\textwidth}
\begin{lstlisting}[captionpos=b, basicstyle=\fontsize{0.27cm}{0.27cm}]  
    pattern EQ_NEQ for
        e: EQ_BIN_OP
    with
        e1, e2: EXPRESSION
    where
        e1 = e.first
        e2 = e.second
    fix
        NEQ_BIN_OP [first $\leftarrow$ e1, second $\leftarrow$ e2]
    end
\end{lstlisting}
\end{minipage}
\end{figure}

%\textcolor{red}{[VIKTORIYA: IF YOU WANT TO ADD AN EXAMPLE OR TWO THAT’S FINE, BUT MAKE SURE THAT THEY ARE CREDIBLE. IF YOU ARE NOT SURE MAYBE BETTER STOP HERE. I’LL LET YOU DECIDE.]}

The examples also illustrate the purpose --- and limits --- of the BUGFIX language. Two arguments may or may not be in the right order; swapping them may or may not be the proper fix. Deciding on these questions is beyond the scope of BUGFIX: it is the task of Automatic Program Repair methods and tools. BUGFIX does not attempt to provide a magical wand for APR, but provides the (brilliant) devisers of magic wands with a way to develop, test, validate, experiment, refine, explain, implement and publicize their contributions, and to compare their success in APR, in objective ways, to the results of magic wands produced by other (brilliant) developers.

\section{Generality assessment} \label{assessment}
The above examples are simple, but the current initial version of BUGFIX, using the ideas just outlined through these examples, make it possible to cover a wide range of common bugs.

To address the issue of generality, our design goal for BUGFIX has not been to address the widest possible range of conceivable bugs, which might lead to an ambitious but unusable contraption; we have taken instead the pragmatic approach of looking at the most common types of bugs; specifically, bugs that are both:
\begin{itemize}
    \item Actual code bugs (rather than high-level design or requirements bugs which, as noted, are harder to handle), with clear applicable fixes.
    \item Often found in actual code, as illustrated --- objectively --- by the empirical study of large software repositories (Linux, Eclipse, Apache).
\end{itemize}

We call such bugs, the ones most conducive to successful APR work, \textbf{Low-Hanging Bugs} (LHBs). 
Fortunately a significant amount of work already exists on the analysis of bugs and fixes for major software repositories \cite{just2014defects4j}. While not complete, it provides a good initial catalog of bugs, from which we started with our own ongoing studies of repositories and used them to collect LHBs. 

%Fortunately a significant amount of work already exists on the analysis of LHBs for major software repositories \cite{just2014defects4j}. While not complete, it provides a good initial catalog of bugs, which we completed with our own ongoing studies of repositories. 

This analysis of bugs through both the literature and our own studies leads to a set of Low-Hanging Bugs that seems to occur widely. 
Specifically, the current analysis uses the Defects4J \cite{just2014defects4j} dataset, which is a widely used benchmark for automatic repair of Java programs. Through an analysis of the 364 representative bugs in the Defects4J dataset, we find 51 LHBs, accounting for 14\% of the total. 
For the rest of the non-LHBs (86\%), their fixes either involve multiple lines of code or cover various types of program constructs, which makes it difficult to generalize fix patterns.
These ever-recurring bugs, identified throughout repositories, appear worthy of particular attention for APR research. The most significant sub-categories are:

\begin{itemize}
    \item Missing Null checking (25.5\% of LHBs)
    \item Incorrect variables (23.5\%)
    \item Bugs related to -/+1 (13.7\%)
    \item Misuse of order operators ($<$, $\leq$, $\geq$, $>$ etc.) (11.7\%)
    \item Misuse of False/True  (7.8\%)
\end{itemize}

%\textcolor{red}{[@VIKTORIYA: PLEASE COMPLETE WITH MOST COMMONLY OCCURRING 5 TO 8 (as you see fit) CATEGORIES, GIVING ONLY – AS I HAVE DONE ABOVE --THE PERCENTAGE FROM LHBs, NOT FROM THE WHOLE SET. ORDERED BY DECREASING PERCENTAGE. IT’S OK TO ROUND FIGURES TO FULL INTEGERS – I LEFT 23.5 SINCE IT’S NOT CLEAR HOW TO ROUND IT.]}

The full list of categories and the current result of our analyses on Defects4J is available in the Github repository\footnote{BUGFIX: github.com/apr-2024/BUGFIX}. 
As ongoing work, we also aim to include the EiffelBase library \cite{meyer1994reusable} in the analysis --- the initial implementation of EiffelBase had a significant number of bugs, many of them non-trivial, and has served as the basis for several earlier studies of bugs and fixes \cite{wei2012branch, 6776507, 7203042}.
We intend to continue to maintain and develop this repository as a general community resource to help APR researchers and product developers.

The potential for developing the BUGFIX language for generality is in principle unbounded: its scope could theoretically encompass any program or design transformation. For BUGFIX to be of practical use, it should strike the right balance between generality and simplicity. We are adopting a cautious approach favoring the second of these criteria, keeping BUGFIX a small language, guided by its applicability to bug and fixes that do appear frequently in practice --- in other words, Low-Hanging-Bugs, where the inspiration comes not from our own intuition or opinions but from the empirical, objective study of credible bug and fix repositories.

\section {Abstract interface} \label {interface}
To guide the specification and further development of BUGFIX, we are relying both on an example concrete syntax illustrated above but, more fundamentally, on an abstract programming interface (API). The original version of the interface is written in Eiffel, since this language offers powerful abstraction mechanisms, particularly the Design by Contract specification facilities; for example the “\e{where}” clause of BUGFIX illustrated by the \e{PLUS_MINUS} example above can directly be expressed by an Eiffel “precondition” (\e{require} clause).  From this basic form of expressing BUGFIX elements (constructs and bug-fix patterns), we will make available others, notably:

\begin{itemize}
    \item The concrete syntax form, as illustrated, for easy human comprehension.
    \item Libraries in other common languages, such as Java or Python (using an open mechanism allowing community contributions).
    \item Non-programming-language forms, for direct consumption by tools, in binary formats or exchange formats such as JSON or XML.
\end{itemize}

The aim is to encourage the development of a wide range of bug and bug-fix patterns, open to contributions by all members of the APR community.

\section{Further applications} \label{applications}
BUGFIX supports the specification of language constructs and bug-fix patterns. These patterns are, more generally, code-transformation schemes. Besides their application to Automatic Program Repair, which the main focus of this article, they have potential uses for other efforts involving predefined schemes for program transformation. Two notable potential examples are:

\begin{itemize}
    \item \textbf{Fault seeding}. A common technique in testing, and more specifically the analysis of test case quality, is to introduce faults (bugs) artificially into programs. For the validity of the corresponding applications, these faults should as much as possible reflect the patterns of actual bugs produced by programmers. To describe fault-seeding schemes, one can use BUGFIX: the fix part describes the correct pattern, and the bug part describes the seeded bug.
    \item \textbf{Verification condition generation}. Software verification, for example in Hoare-style axiomatic semantics, needs assertions about the program, for example loop invariants. A modern, effective software verification environment should help programmers produce such “verification conditions”, facilitating a task which can be conceptually hard and, even when it is not, remains tedious and time-consuming. The Daikon tool \cite{ernst2007daikon} has pioneered this line of research by proposing invariant patterns. Such patterns can be described in BUGFIX (ignoring in this case the “fix” part of specifications).
\end{itemize}

More generally, program transformation is a recurring need of software development tools, arising in many different applications; BUGFIX can provide a general program-transformation specification framework.

\section{Related work} \label{related_work}

The idea of identifying and/or analyzing common bug or fix patterns is not new, but it has been mostly applied in specific contexts or programming languages: the analysis of bug and fix patterns \cite{campos2017common, soto2016deeper} focuses on the patterns for Java program; 
the work presented in \cite{livshits2005dynamine} and \cite{williams2005automatic} identified project-specific bug patterns with focuses on method call pairs (two methods should be called in pairs) or checking return value. 
Sun et al. \cite{sun2017empirical} identified different categories of bugs and commonly used fix patterns in Machine Learning projects.
In contrast to the above work, the approach presented in this paper aims to provide a unified framework that allows to derive bug and fix patterns in a more general context, involving various languages or application domains.

Duraes et al. \cite{duraes2006emulation} presented a classification of defects, which is an extension of the Orthogonal Defect Classification (ODC) \cite{chillarege1992orthogonal}, with the main focus on the Emulation of Software Faults.
Catolino et al. \cite{catolino2019not} presented an taxonomy of bug root causes. The primary motivation of their work, however, lies in expediting error resolution for developers, not identifying commonly used bug and fix patterns.

A repository described in \cite{lin2017quixbugs} contains a same set of bugs in two different languages (Java and Python); it provides a benchmark for evaluation of multilingual APR tools. BUGFIX shares the similar vision -- to serve as a common platform that allows evaluation and comparison of APR techniques involving different programming languages.

%In \cite{sobreira2018dissection}, nine repair patterns were identified. Each pattern has an example. The authors analyzed a dataset containing at that time 395 real bugs from open source code, namely Defects4J \cite{just2014defects4j}. Currently the dataset contains 835 bugs (plus 29 deprecated bugs). 

\section{Limitations, conclusions and future work} \label{future_plans}
This workshop contribution describes the inception and first results of a project intended to provide the Automatic Program Repair with a common frame of reference, BUGFIX, including a conceptual framework, a language in the complete sense of the term (abstract syntax, API, concrete syntax, hooks to an extendible set of programming languages), and a rich and reliable repository of important and representative examples of bugs and possible fixes, allowing them (as noted at the end of section 2) to apply their insights and artifacts to standard examples and validate their effectiveness against other work.

The limitations of the current state of the work have been made clear throughout this article. The most obvious is its work-in-progress nature. It is important, however, to note that the key design decisions have been made (as described in the preceding sections) and are here to stay. They follow from a careful examination of key criteria and an attempt at obtaining an effective tradeoff; the criteria are: simplicity; learnability; generality; practicality (based on a long-running analysis of actual bugs as actual people produce them, from the empirical analysis of actual software repositories). They determine the essential nature of the BUGFIX framework and are the principal contribution of the work.

The self-admitted tentative nature of the present description is not just, however, an item in its list of limitations. It is also, more importantly, one of its intentional features. This article is a submission to a workshop and has been designed that way. Instead of a solution claiming to be fully finalized, which might fit the authors’ personal Weltanschauung of software engineering but fail to miss the real needs of many developers, it proposes an initial version of BUGFIX, sufficiently explicit and developed to show the essential insights, contributions and applications, but still open to refinement. By using this approach, we expect to elicit constructive feedback from the APR community and steer the future development of BUGFIX towards a final result that will be the most useful possible to a broad set of APR innovators.

Already in its current state as specified in the preceding sections, BUGFIX provides a general framework for describing and processing bugs and their fixes. We hope that its introduction will --- even in a modest way --- help advance one of the key goals towards  the more general quest by the software engineering profession to provide the world with better software: the goal of equipping software developers --- when they occasionally let bugs slip into their programs, as almost all of them with the possible exception of Donald Knuth inevitably do, and will continue to do for a long, long time ---  with effective, useful and correct suggestions of Automatic Program Repair.

\bibliographystyle{ACM-Reference-Format}
\bibliography{reference}

\end{document}